\documentclass[iop]{emulateapj}
\usepackage{apjfonts}
\submitted{Accepted to ApJ}
\shorttitle{Dusty Wind-Blown Bubbles}
\shortauthors{Everett \& Churchwell}
\newcommand{\um}{{\,\mu\rm m}}

\begin{document}

\bibliographystyle{apj}

\title{Dusty Wind-Blown Bubbles}

\author{by J.E. Everett\altaffilmark{1,2,3}, 
  E. Churchwell\altaffilmark{1}}
\altaffiltext{1}{University of Wisconsin--Madison, Department of Astronomy}
\altaffiltext{2}{University of Wisconsin--Madison, Department of Physics}
\altaffiltext{3}{Center for Magnetic Self-Organization in Laboratory and Astrophysical Plasmas}
\email{everett@physics.wisc.edu}

\begin{abstract}
Spurred by recent observations of 24$\micron$ emission within
wind-blown bubbles, we study the role that dust can play in such
environments, and build an approximate model of a particular
wind-blown bubble, `N49.'  First, we model the observations with a
dusty wind-blown bubble, and then ask whether dust could survive
within N49 to its present age (estimated to be $5 \times 10^5$ to
$10^6$ years).  We find that dust sputtering and especially dust-gas
friction would imply relatively short timescales ($t \sim 10^4$~years)
for dust survival in the wind-shocked region of the bubble.  To
explain the $24\micron$ emission, we postulate that the grains are
replenished within the wind-blown bubble by destruction of embedded,
dense cloudlets of ISM gas that have been over-run by the expanding
wind-blown bubble.  We calculate the ablation timescales for cloudlets
within N49 and find approximate parameters for the embedded cloudlets
that can replenish the dust; the parameters for the cloudlets are
roughly similar to those observed in other nebula.  Such dust will
have an important effect on the bubble: including simple dust cooling
in a wind-blown bubble model for N49, we find that the luminosity is
higher by approximately a factor of six at a bubble age of about
$10^4$ years.  At ages of $10^7$ years, the energy contained in the
bubble is lower by about a factor of eight if dust is included; if
dust must be replenished within the bubble, the associated
accompanying gas mass will also be very important to wind-blown bubble
cooling and evolution.  While more detailed models are certainly
called for, this work illustrates the possible strong importance of
dust in wind-blown bubbles, and is a first step toward models of
dusty, wind-blown bubbles.
\end{abstract}


\section{Introduction}\label{Intro}

In addition to the energy, mass, and momentum input provided by the
deaths of massive stars in supernovae, early-type stars also launch
hypersonic ($v \sim 1000 - 3000$\,km\,s$^{-1}$, whereas $c_{\rm s,ISM}
\sim 10$\,km\,s$^{-1}$) winds with mass outflow rates of order
$10^{-7}$ to $10^{-6}$\,M$_{\odot}$\,yr$^{-1}$ \citep[although
clumping within the wind may imply smaller mass-outflow rates;
see][]{Hillier2009}.  Such winds drive forward shocks both into the
surrounding interstellar medium and reverse shocks back into the
stellar wind; these reverse shocks then heat the stellar wind into a
pressurized bubble surrounding the young star, resulting in objects
known as wind-blown bubbles \citep[WBBs; for reviews see,
e.g.,][]{ZhekovMyasnikov2000, Chu2003, Arthur2007, Chu2008}.  WBBs are
therefore thought to have four major components: (1) an inner,
free-flowing, hypersonic wind surrounded by (2) the hot, post-shocked
wind plasma, which is enclosed by (3) a thin shell of $10^4$\,K gas
which resides in (4) the ambient interstellar medium \citep[see Figure
1 of ][for a sketch of these components; they also form the basis of
our Fig.~\ref{n49Schematic}]{WeaverEtAl1977}.

The structure of these WBBs is useful to study in that they help us
understand the transformation of gas phases in the interstellar
medium, and more generally, the impact of massive stars on their host
molecular clouds, the heating of ISM gas, and the input of energy into
turbulence in the ISM \citep[e.g.,][]{Hensler2008}; they may even
trigger star formation on their periphery
\citep[][]{ChurchwellEtAl2006}.  Observations of WBBs also place
constraints on stellar-wind power over the evolution of early-type
stars.  Finally, WBBs are important for setting the stage for the
star's subsequent supernova explosion or perhaps gamma-ray burst.

These wind-blown bubbles and their impact on the interstellar medium
have been observed for many years \citep[e.g.,][]{JohnsonHogg1965,
Smith1967, ChuLasker1980, NazeEtAl2001, ChurchwellEtAl2006,
WatsonEtAl2008}.  Theoretical work in understanding these observations
has been pursued in parallel by many researchers
\citep[e.g.,][]{JohnsonHogg1965, Mathews66, Pikelner68, Dyson73,
CastorEtAl75, Falle75, WeaverEtAl1977, HartquistEtAl1986,
ArthurEtAl1993, GarciaSeguraMacLow1995, GarciaSeguraEtAl1996,
ArthurEtAl1996, PittardEtAl2001a, PittardEtAl2001b, NazeEtAl2002,
FreyerEtAl03, FreyerEtAl2006}.  These dynamical studies of wind-blown
bubbles have included detailed modelling of the effect of a
surrounding inhomogeneous medium \citep[and especially, the role of
clumps in the surrounding medium, which will be become important later
in this work; see][for example]{ArthurEtAl1993, PittardEtAl2001b,
NazeEtAl2002}.  To the present day, however, such models have not
included the role of dust.  Recent observations by
\citet{ChurchwellEtAl2006} and \citet{WatsonEtAl2008} present evidence
that dust exists within WBBs.  What role does this dust play in the
structure and the evolution of the bubbles, and how might dust modify
the impact of WBBs on the surrounding interstellar medium?

This paper aims to study the role of dust in wind-blown bubbles in the
context of observations of one WBB: N49.  We will start by briefly
outlining the observations of N49 in Section~\ref{bubbleObs}.  This paper
will then examine the role of dust within N49; in particular, we ask
if dust can explain the observed $24\um$ emission around N49 (examined
in Section~\ref{modelDef}), and then ask if dust can survive within the
$10^6$-$10^7$\,K, post-shocked wind-blown bubble in N49 in
Section~\ref{dustSurvival}.  Then, in Section~\ref{proposedModel}, we offer a
simple model to explain the observations and our results on dust
survival.  Our conclusions are presented in Section~\ref{conclusions}.

\section{Observed Mid-IR Continuum Distribution}\label{bubbleObs}

This paper considers the mid-infrared signature of one particular WBB:
`N49.'  N49 was discovered in the infrared and cataloged by
\citet{ChurchwellEtAl2006} as part of a manual search of the Galactic
Legacy Infrared Mid-Plane Survey Extraordinaire \citep[GLIMPSE;
  see][]{Benjamin2003, ChurchwellEtAl2009} taken by the
\textit{Spitzer Space Telescope} \citep{WernerEtAl2004}.
\citet{ChurchwellEtAl2006} found over 300 partial and closed rings in
their search of the GLIMPSE data; their paper presents an analysis of
the characteristics of the (incomplete) large population of bubbles
reported.  A later catalog for bubbles within $|l| < 10^\circ$ in the
Galaxy found a comparable number.  These bubbles were identified by
rings of $8\um$ emission that were found to enclose $24\um$ emission;
a few such infrared bubbles had been detected before by the
\textit{Infrared Space Observatory} and by the \textit{Midcourse Space
  Experiment} \citep[e.g.,][]{PasqualiEtAl2002, DeharvengEtAl2005,
  VasquezEtAl2009}.

N49 was later the focus of a more detailed study by
\citet{WatsonEtAl2008}.  For that study, N49 was chosen because of its
relatively symmetric appearance: for instance, the $8\um$ emission was
very well fit by assuming a spherical shell \citep[see Fig.~12
of][]{WatsonEtAl2008}.  This implies that perhaps N49 is a useful test
case for comparing with models of WBB evolution.
\citet{WatsonEtAl2008} also examined $24\um$ emission observed with
the Multiband Imaging Photometer aboard \textit{Spitzer}
\citep[MIPS;][]{RiekeEtAl2004} as part of the the MIPS Galactic Plane
Survey \citep[MIPSGAL; see][]{CareyEtAl2009}, comparing that data to
$8\um$ \textit{Spitzer} observations and to 20\,cm radio observations
\citep{HelfandEtAl2006}.  These data showed that emission at $24\um$
and 20\,cm are coincident, sharing the same annular geometry (although
the 20~cm data, indicating free-free emission from ionized gas, does
not drop off quite as quickly with radius as the $24\um$ emission);
both were also enclosed by the larger-radius $8\um$ emission of the
photodissociation region (PDR) shell, strongly indicating that the
dusty post-shocked gas resides within the outlying $8\um$ PDR shell.
However, it is not at all clear if a reasonable mass of dust could
explain the observed emission, and if dust could survive in that
post-shocked gas environment.  And, if dust can exist within the
bubble, what role might such dust play in the evolution of wind-blown
bubbles?  These are the questions that the present paper sets out to
answer.

\section{Modelling the Mid-IR Continuum Distribution}\label{modelDef}

Our first task was to see whether dust emission (alone) is a
reasonable fit to the observations.  To synthesize observations of
dust in WBBs, we have run photoionization simulations using `Cloudy'
\citep[version 07.02.01, last described by][]{FerlandEtAl98}.  In
those Cloudy simulations, we investigated emission from a
constant-temperature ISM plasma with an included ``normal'' dust grain
population \citep[using the dust-grain size distribution
of][]{MathisEtAl1977} with a dust-mass-to-gas-mass ratio of
$6.4\times10^{-3}$.  We employed Cloudy\_3D \citep{Morisset06} to map
the 1D Cloudy simulations into a spherically-symmetric 3D bubble that
was then projected onto the plane of the sky to compare with
observations.  Note that Cloudy cannot model the dynamic post-shock
structure of the WBB, so for photoionization studies, we model the
impact of dusty gas on a \textit{static} bubble of uniform temperature
which we set.  At this early stage, we have no dynamical model of dust
fluxes into, or out of, the bubble; we are examining the emission from
a dusty ISM plasma within the bubble, with a standard ISM dust-to-gas
ratio.

The parameters for the simulations and analysis are summarized in
Table~\ref{cloudyModelParams}, along with references for the values
used.  We describe and justify those parameter choices in the
paragraphs that follow.

\begin{table}[h]
\begin{center}
\begin{tabular}{|l|l|l|}
\hline \hline
\emph{Parameter} & \emph{Value} & \emph{Reference} \\
\hline
central star & O5V & \citet{WatsonEtAl2008}\\
$\dot{M}_{\rm wind}$ & $1.5 \times 10^{-6}~M_{\odot}\,{\rm yr}^{-1}$  & \citet{VinkEtAl2001}\\
%& $M_{\rm stellar} = 60$~$M_{\odot}$& \\
%& $T_{\rm stellar} = 3.9\times10^4$~K & \\
%& $L_{\rm stellar} = 10^{5.6}$~$L_{\odot}$& \\
%& $Z = Z_{\odot}$& \\
distance & $5.7$~kpc & \citet{ChurchwellEtAl2006} \\
age & $\sim10^6$~yrs & \citet{WatsonEtAl2008} \\
$<n_{\rm inner,~dusty~gas}>$ & $19.1$~cm$^{-3}$ & (Best-Fit, 24\,$\mu$m
data) \\ % 19.125 cm^-3 is the best fit number
gas \& dust density &  &\\
\hspace{1.5em} power law: $r^{\alpha}$  & $\alpha$ = -2 & \textit{Assumed}  \\
$r_{\rm inner}$ & $0.6$~pc & (Best-Fit, 24\,$\mu$m data) \\ % log(r0)
				% = 18.24 is the real best-fit number
$r_{\rm outer}$ & $2.2$~pc & interface of PDR shell \\ 
dust & ISM grain & \citet{MathisEtAl1977}\\
 & \hspace{0.2cm}distribution & \hspace{0.2cm}\citet{vanHoofEtAl2004}\\
$T_{\rm e, bubble}$ & $3.5 \times 10^6$~K & Our dynamical model
\\ 
 & & \hspace{0.2cm}(Section~\ref{modelResults}) \\
\hline
\end{tabular}
\end{center}
\caption{The values used for parameters in the Cloudy simulations of
 the dusty wind-blown bubble, N49.} \label{cloudyModelParams}
\end{table}

For the central star and its central stellar continuum, we include a
star of class O5V \citep{WatsonEtAl2008}.  Using Cloudy's `wmbasic'
stellar continuum models \citep[][which include line-blanketing by
stellar winds]{PauldrachEtAl2001}, and calling on the models in
Table~1 of \citet{MartinsEtAl2005}, we set the star's luminosity to
$10^{5.51}~L_{\odot}$, the surface temperature to $T_{\rm eff} =
41540$~K, the surface gravity to $\log(g) = 3.92$, and the star's
metallicity to solar.  These parameters correspond to a
$37.3~M_{\odot}$ star, with a flux of hydrogen-ionizing photons of
$\log q_0 = 10^{24.4}$~photons~cm$^{-2}$~s$^{-1}$ or $\log Q_0 =
10^{49.3}$~photons~s$^{-1}$ into $4\pi$ steradians, very close to the
value cited by \citet{WatsonEtAl2008}.

The stellar mass loss rate, $\dot{M}_{\rm wind}$ is not used directly
in the Cloudy simulations, but used in our analysis of the velocity of
the post-shocked gas to estimate (later) the dust-gas friction that
drags dust out of the post-shocked wind region.  For the stellar
parameters we've adopted, the IDL code `Mdot.pro' from
\citet{VinkEtAl2001} yields a mass outflow rate of
$1.4\times10^{-6}$~M$_{\odot}$~yr$^{-1}$ and $v_{\infty} \sim
2600$~km~s$^{-1}$.  This yields a wind kinetic luminosity of $3 \times
10^{36}$~ergs~s$^{-1}$.

The kinematic distance to N49 is $5.7 \pm 0.6$~kpc
\citep{ChurchwellEtAl2006}.  Given this distance, the size of the
bubble, $r_{\rm outer}$, is set by the inner radius of PAH emission,
so is observationally estimated \citep{WatsonEtAl2008} as 2.2\,pc.
The density of ambient gas surrounding N49 is not well constrained,
but unpublished observations of ${\rm NH}_3$ emission (Cyganowski et
al., in preparation) surrounding N49 indicate a high ambient gas
density of order $10^4$\,cm$^{-3}$.  The age of N49 is not well
constrained, but if the external medium has a gas density of order
$10^4$~cm$^{-3}$, the age of a wind-blown bubble of the inferred size
for N49 is approximately $5 \times 10^5$ to $10^6$~years
\citep{WatsonEtAl2008}.

Again, since Cloudy does not simulate the heating of the wind's
post-shocked gas, we start by modelling the WBB with a uniform
temperature for the post-shocked gas.  For most of the simulations, we
have chosen $3.5\times10^6$~K as that constant temperature; this value
for the temperature is informed by our own simple dynamical models of
wind-blown bubble dynamics with dust cooling (see
Section~\ref{modelResults}).  Changes in the WBB temperature affect the
amount of dust that is inferred; increasing the temperature to
$10^7$\,K as suggested in the dust-free models of \citet{FreyerEtAl03}
would reduce the amount of dust required, but not eliminate it.

Finally, we assume that the grain distribution, which we specify in
Cloudy, is an ISM-type grain distribution \citep{MathisEtAl1977,
vanHoofEtAl2004}.  We explain below, in Section~\ref{cloudyModelFits}
why that grain distribution seems to be preferred by the observations.

\subsection{Fitting 24$\um$ observations with Cloudy models}\label{cloudyModelFits}

\begin{figure}[ht!]
\begin{center}
\includegraphics[width=6cm,angle=90]{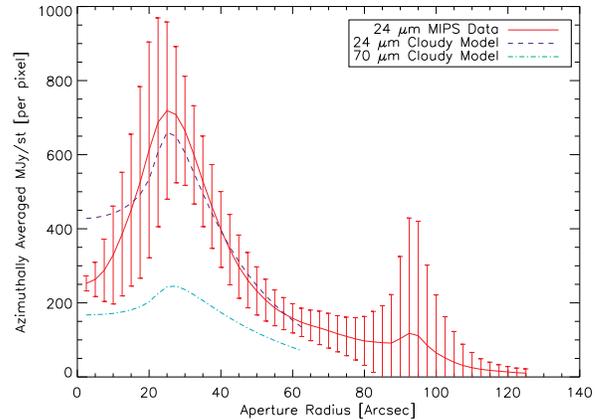}
\caption{The best-fit model for 24$\um$ emission within the wind-blown
  bubble, using an ISM dust-grain distribution with the
  photoionization model parameters given in
  Table~\ref{cloudyModelParams}.  The solid line shows the azimuthally
  averaged flux from \textit{Spitzer} MIPS data, while the dashed line
  shows the photoionization results, using the same azimuthal
  averaging.  The error bars show the $1\sigma$ variation in the flux
  around the averaged aperture.\label{24micronEmissionComparison}}
\end{center}
\end{figure}

First, we ask whether dust emission can reproduce the \textit{Spitzer}
24$\um$ observations of N49.  Using the above-defined Cloudy and
Cloudy\_3D models, we simulate the emission in the \textit{Spitzer}
MIPS 24$\um$
band\footnote{http://ssc.spitzer.caltech.edu/mips/descrip.html} by
defining a filter centered on 23.7$\um$ that spans from 21.65$\um$ to
26.35$\um$; we also define, for later use, the MIPS 70$\um$ band,
centered on 71.0$\um$, spanning wavelengths from 61.5$\um$ to
80.5$\um$.  To compare to the data, we take azimuthal averages of both
the \textit{Spitzer} data \citep{ChurchwellEtAl2006, WatsonEtAl2008}
and the photoionization model.  For the \textit{Spitzer} MIPS data, we
take the flux average over 50 concentric annuli that have a $\Delta r$
of 2 pixels (with one pixel = 1.25''), centered on the centroid of
flux within the bubble; the result of this azimuthal averaging is
shown by the solid line in Figure~\ref{24micronEmissionComparison}.
The background emission is estimated by calculating the average flux
in an annulus with an inner radius of 130 pixels from the bubble
centroid, and an outer radius of 150 pixels from the bubble centroid,
where there are no background stars.  The $1\sigma$ variance in the
per-pixel flux is shown by the error bars in the figure; plotting this
variance helps to confirm that the rise (at about 25'') and fall of
the $24\um$ emission is present, but with some variations that we have
averaged out.

We first performed a large-scale, manual search of parameter space for
the approximate gas density and the approximate WBB inner radius that
reproduced the observations; we started our search at $n = 1\,{\rm
  cm}^{-3}$ and $r_{\rm inner} = 0.5$\,pc (the last value is taken
from the approximate observed inner radius of $24\um$ dust emission).
After manually increasing the gas density, $n$, to reproduce
approximately the observed dust emission, we then set up an automated
grid search of the parameter space stretching from $\log(r_{\rm
  inner}/{\rm cm})$ = 18.15 to 18.30 in steps of 0.05, and with
$<n_{\rm dusty~gas}(r_{\rm inner})>$ = 19.00 to 21.00 cm$^{-3}$ in
steps of 0.25, followed by higher-resolution searches with step-sizes
down to 0.01 in $\log(r_{\rm inner})$ and 0.0625 in $<n_{\rm
  dusty~gas}(r_{\rm inner})>$.  This allowed much more accurate
estimates of the WBB parameters.  In all of our parameter searches,
there was no apparent degeneracy in $\chi^2$ for the two parameters
investigated, and only one minimum in $\chi^2$ was found, producing a
unique fit to the data (we used Poisson errors in calculating
$\chi^2$, weighting the large-scale radial decline in $24\um$ emission
with radius, and not the central depression in emission).

We find a reasonable match to the 24$\um$ emission, as shown by the
dashed line in Figure~\ref{24micronEmissionComparison}.  For this fit,
we specify that the dusty gas density scales as $n(r) \propto r^{-2}$
and find the best-fit initial density and inner radius (again, see
Table~\ref{cloudyModelParams} for input parameters and resultant
best-fit parameters).  This particular power-law for the density was
chosen because it fit the observed pattern of 24$\um$ emission with
radius; we will offer a theoretical explanation of the dropoff in
density with a more detailed model in Section~\ref{proposedModel}.

Note that while we include a central hole in the dust distribution,
the model does not reproduce the strength of 24$\um$ emission towards
the center of N49.  This may perhaps be the result of some
non-sphericity in the N49 bubble, resulting in less dust along the
line of sight to the central source.  Or, as we suggest later for
other reasons, if the dusty medium inside the WBB is clumpy, that
could result in an enhanced contrast between the central hole and
shell of 24$\um$ emission for some lines of sight (K. Wood 2009,
private communication).

This fit shows that it is possible, with an ISM-like dust distribution
and with a plausible gas density \& temperature, to reproduce the
$24\um$ observations.  But is an ISM-like dust distribution the only
one that is consistent?  We can check further whether our dust
distribution is reasonable by comparing to observations at other
wavelengths; in particular, \textit{Spitzer} MIPS data at $70\um$
taken simultaneously with the $24\um$ observations.  If small dust
grains are preferentially destroyed, then N49 would have a dust-grain
distribution weighted towards large-dust grains; requiring such large
grains to reproduce the $24\um$ emission would require a large
increase in the total number of large grains; that increase would
produce a greater amount of $70\um$ emission relative to $24\um$
emission than for a normal ISM-like dust distribution.

It is important to point out, however, that the observed $70\um$
emission has many contributions: it is a function of the total
starlight intensity in the vicinity of N49, including stars other than
the central OV star, and is not due to the hot post-shock gas in the
WBB, as for the $24\um$ emission.  It is therefore not surprising that
(as we will see) there is a significant background of $70\um$ emission
observed far from the post-shocked region of N49.  In addition, the
currently available $70\um$ \textit{Spitzer} images have obvious
stripping artifacts (i.e., not a background matching problem), one of
which runs directly through the image of the N49 bubble.  However, to
our knowledge, the $70\um$ data have been carefully calibrated and
should provide a reasonably correct approximate upper limit to
constrain the dominance of large dust grains in N49.

So, to examine this constraint, we ask whether a grain population with
only large grains could also fit the data.  We test this idea by
finding the best-fit model to the \textit{Spitzer} $24\um$ MIPS data
with a large-grain-only population (defined as only having grains of
size 0.1 to 1$\um$).  For this fit, the predicted $24\um$ emission is
very similar to that shown in Figure~\ref{24micronEmissionComparison}:
the best-fit initial radius for this model is $\log(r_{\rm inner}/{\rm
cm})$ = 18.3 and $<n_{\rm dusty~gas}(r_{\rm inner})> =
17.4$\,cm$^{-3}$.  Most importantly, though, to fit the $24\um$ data,
the dust/gas ratio is $10^{4.8}$ times greater than the normal ISM
dust-grain case, so we are clearly forced to an \textit{extreme} case
of a large-dust-grain dominated bubble to fit the $24\um$ emission in
a similar fashion as shown in Figure~\ref{24micronEmissionComparison}.

\begin{figure}[h!]
\begin{center}
\includegraphics[width=6cm,angle=90]{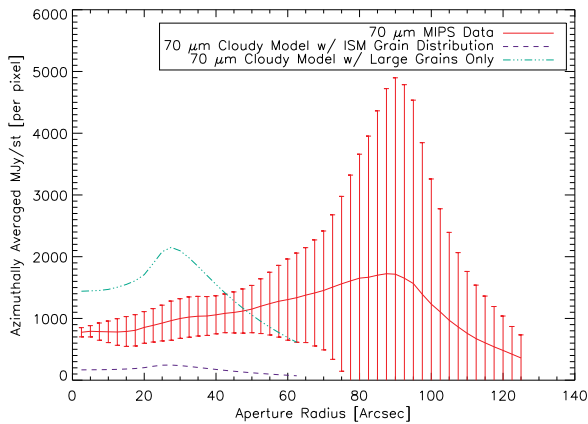}
\caption{As in Figure~\ref{70micronEmissionComparison}, except that
  data and predictions are at $70\um$, and we include a prediction for
  the case where dust grains are limited to have sizes of 0.1 to
  1$\um$ only. Such a large-grain population results in very high
  70$\um$ flux compared to the observed flux.  The error bars show the
  $1\sigma$ variation in the flux around the averaged aperture.
  \label{70micronEmissionComparison}}
\end{center}
\end{figure}

Figure~\ref{70micronEmissionComparison} compares the observed $70\um$
emission with the predictions from Cloudy models with a normal
ISM-like dust-grain distribution (shown with the dashed line, as in
Fig.~\ref{24micronEmissionComparison}) and the large-grains-only
distribution (shown with the dot-dashed line).  The WBB with large
dust grains only clearly overpredicts the observed emission (by more
than a factor of two at the peak of the predicted emission), whereas a
WBB model with an ISM-like dust-grain distribution is, at its peak, a
factor of four below the observed emission.  In addition, the observed
$70\um$ emission peaks much further from N49's central star than the
$24\um$ emission, reaching a maximum very near young stars that appear
to be situated on N49's rim.

Returning to the comparison of the fits, this comparison shows that,
along with the extreme over-abundance of dust required to fit the
$24\um$ emission with a large-grain-only dust distribution, the
observations seem to require a significant population of small dust
grains as well.  Again, we caution that the $70\um$ emission contains
contributions from all surrounding stellar sources, so we do not
expect the Cloudy model to match the $70\um$ emission, but it is
clearly a valuable constraint on the dust population in N49, and other
wind-blown bubbles (Watson et al., in preparation).

\begin{figure}[t!]
\begin{center}
\includegraphics[width=6cm,angle=-90]{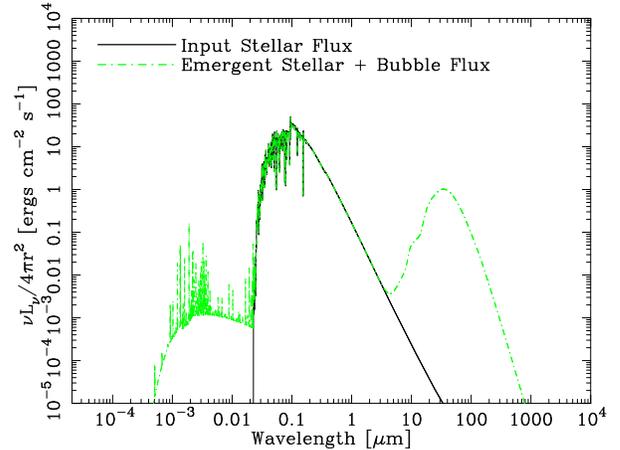}
\caption{A comparison of the continua from the best-fit
  photoionization model.  The solid line shows the stellar continuum
  that is input into Cloudy.  The green, dashed line shows the full
  continuum emerging from the star+wind-blown bubble system.  In the
  X-ray and UV (from $10^{-4}$ to $10^{-2}\,\um$), the $3.5 \times
  10^6$\,K gas of the wind-blown bubble dominates, while at $\lambda
  \ga 4\um$, dust emission clearly
  dominates.\label{continuumComparison}}
\end{center}
\end{figure}

Let us now examine where the emergent 24$\um$ flux is coming from: is
it indeed dominated by dust, and how does the dust flux compare with
the stellar flux and other contributions?  To examine this, we show
the model continua in Figure~\ref{continuumComparison}.  The stellar
flux is shown by the solid line (which shows the input stellar flux);
it is reasonably approximated with a blackbody, although stellar-wind
absorption features (around $\sim 0.1\um$) are present.  The green,
dashed line shows the flux that emerges from the dusty, wind-blown
bubble and star, together.  At wavelengths from 10$^{-4}$ to 10$^{-2}$
$\um$ (in the UV to X-ray regime of approximately 1 to 100
$\mathring{{\rm A}}$), the flux is dominated by the hot wind-shocked
gas in the bubble, where we have set $T = 3.5 \times 10^6$~K.  On the
other side of the stellar continuum, at long wavelengths of $\sim
4\,\um$ and beyond, the dust emission dominates the continuum.  In
fact, one can see the $10\um$ silicate feature superposed on the dust
continuum.  We have checked that no other line features contribute in
this region, in our model; of the atomic lines in Cloudy, \textsc{H I}
recombination lines in the \textit{Spitzer} $24\um$ bandpass
contribute more than any other atomic lines, and they remain at a
luminosity nine orders of magnitude smaller than the dust emission.
In this dusty WBB, dust emission dominates in the $24\um$
\textit{Spitzer} bandpass.

For later comparison with dynamical models, an estimate is needed for
the amount of dusty gas required in the wind-blown bubble to reproduce
the observations.  Given the parameters for the best-fit model, the
total mass of dusty gas would be $4.8 \times 10^{33}$~g, or
approximately 2.4\,M$_{\odot}$.  How does that compare with the mass
available in the area swept up by the bubble?  For the same outer
radius, assuming an ambient density of $10^4$~cm$^{-3}$, approximately
$10^4$ solar masses would be available.  Thus, the dusty gas required
is a small fraction of the gas available.

\section{Can Dust Survive in this Environment?}\label{dustSurvival}

These observations and models present a strong case for N49 as a
dusty, wind-blown bubble.  However, there are a variety of processes
that could either destroy or remove dust from the interior of a
wind-blown bubble: sublimation, sputtering, radiative acceleration,
and/or gas-dust friction.  Can dust survive within N49?

\begin{figure}[h!]
\begin{center}
\includegraphics[width=6cm,angle=90]{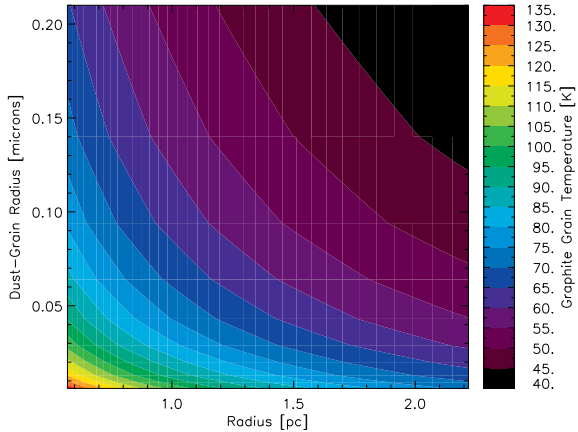}
\caption{The graphite-grain temperature as a function of radius in the
  wind-blown bubble and as a function of dust-grain size.  The grain
  temperatures are significantly lower than the grain sublimation
  temperature of $1500\,K$, so even the smallest, hottest grains do
  not sublimate away. \label{dustGrainTemps}}
\end{center}
\end{figure}

We first examine the possibility of dust sublimation in
Figure~\ref{dustGrainTemps}. In our Cloudy models, the graphite-grain
temperatures are quite low, spanning the range of 40 to 140\,K
depending on position in the wind-blown bubble and on grain-size (the
silicate-grain temperatures are quite similar: the smallest silicate
grains are approximately 25\% cooler at the bubble innermost radius,
and have temperatures within 10\% of the graphite grains over most of
the wind-blown bubble).  For a given grain size, the variation in
grain temperature is only about 50\% throughout the entire bubble.
Overall, the grain temperatures are much lower than the grain
sublimation temperature of $\sim 1500$\,K, so the grains are in little
danger of being sublimated away.  We note that, also, the dust
temperatures given here are quite similar to those observed in other
wind-blown bubbles (Watson et al., in preparation).

\begin{figure}[h!]
\begin{center}
\includegraphics[width=6cm,angle=90]{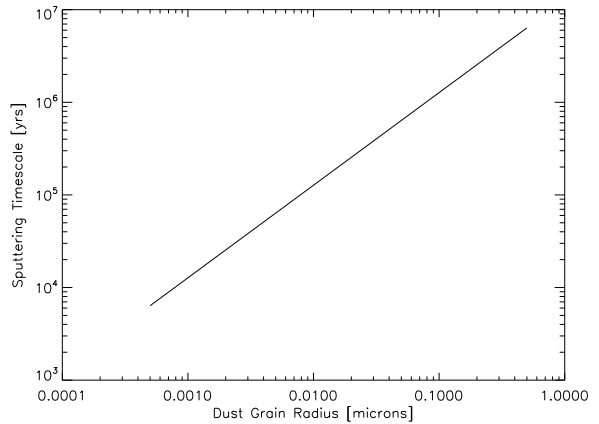}
\caption{The sputtering rates for grains of a variety of sizes within
  the $3.5 \times 10^6$\,K post-shock bubble of N49 for the assumed
  post-shock gas density of $19.1$\,cm$^{-3}$ (the sputtering rate
  increases linearly with gas density).  The sputtering times for
  grains are smaller than the estimated age of the N49 bubble for
  grains of size less than 0.1~$\mu$m.
  \label{dustGrainSputtering}}
\end{center}
\end{figure}

We next test whether the grains are sputtered away by gas-grain
collisions.  For this, we first assume that the grains are stationary
relative to the bulk motion of the post-shock gas (i.e., that the gas
and dust have the same flow velocities), an approximation that we will
later see is quite good, given the dust-gas coupling in the
post-shock region.  We calculate the sputtering rates from the formula
derived by \citet{TielensEtAl94}, and present the results in
Figure~\ref{dustGrainSputtering}.

The sputtering rates depend only on the size of the grain and (for the
assumption that the gas and grains are well-coupled and moving at the
same velocity) the temperature of the surrounding gas.  We can see
that for the assumed temperature of $3.5 \times 10^6$\,K, the
sputtering times for grains with radii less than about 0.1~$\mu$m are
$\la 10^6$\,years, the approximate estimated age of N49 ($\sim
10^6$\,yrs).  We caution that these sputtering timescales should be
recognized as upper limits on the sputtering time; before dust is
accelerated to the gas velocity, the dust grains will suffer
collisions with gas ions of greater relative velocity, and so will
have a somewhat lower survival time than shown here.

Of course, this result is strongly dependent on the gas density in the
post-shock region.  Integrating the equations for a simple (dust free)
\citet{WeaverEtAl1977} wind-blown bubble, we find that for an external
density of $10^4$\,cm$^{-3}$, the bubble's internal density is
approximately $10$\,cm$^{-3}$ at an age of approximately $5 \times
10^5$~years (we will discuss this model and more complicated models
later in Section~\ref{proposedModel}).  This results in much faster
sputtering than in the internal density in the \citet{WeaverEtAl1977}
models, where the internal density was approximately 0.03~cm$^{-3}$,
because of their lower external density of 1~cm$^{-3}$ (at t = $10^6$
years in their Figure 3).

\begin{figure}[h!]
\begin{center}
\includegraphics[width=6cm,angle=90]{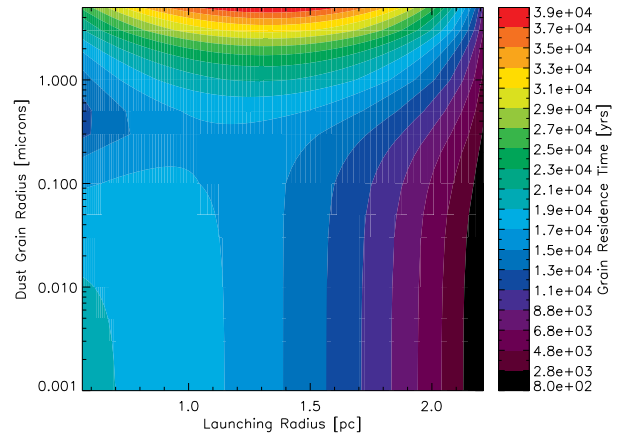}
\caption{The residence time of dust within the wind-blown bubble,
  computed using the frictional force the dust grains feel because of
  the post-shock gas streaming past them.  We calculate the time it
  takes each dust grain to be acceleration from its starting position
  (``launching radius'') to the edge of the nebula, and plot that time
  for a range of dust-grain sizes.\label{gasFrictionAcceleration}}
\end{center}
\end{figure}

We also check how quickly the grains are evacuated from the bubble by
being swept up in the post-shock gas; the results of this calculation
are shown in Figure~\ref{gasFrictionAcceleration}.

To calculate the acceleration of dust due to dust-gas friction, we
make the following assumptions.  First, we assume that the velocity of
the post-shock gas follows the velocity law derived by \citet[][page
379]{WeaverEtAl1977}.  As mentioned previously, we assume a wind
velocity of 2600~km~s$^{-1}$.  We calculate the friction between dust
and gas using the results of \citet{BerruyerFrisch83}, and model the
acceleration of the dust in the presence of that drag force and the
gravitational potential of the central star (we find radiative
acceleration only leads to a small drift velocity between the dust and
gas that ranges between 5 and $\la$15\,km\,s$^{-1}$, depending on
grain size, which is negligible compared to the $\ga
500$\,km\,s$^{-1}$ velocities of the post-shock gas and dust, as shown
below).  We separately derive the acceleration vs. grain size for dust
at a range of initial radii within the wind-blown bubble, and then
calculate how long it takes a dust grain to transit from its starting
position to the edge of the nebula.  This gives us the grain residence
time, which is plotted for a range of launching radii and dust-grain
radii in Figure~\ref{gasFrictionAcceleration}.

As expected, the larger dust grains have greater inertia, and so take
longer to be accelerated by dust-gas collisions, and have
correspondingly longer residence times, but only by a factor of
approximately two.  Again, given the approximate age of N49 at $5
\times 10^5$ to $10^6$\,years, it appears that all dust would be
evacuated from the bubble on timescales of order $10^4$\,years, much
less than the bubble age.  In fact, gas-grain friction would evacuate
dust from the bubble on shorter timescales than the sputtering
timescale; the sputtering timescale is smaller only for very small
grains with sizes of order $0.001\um$.  So, both processes indicate a
much shorter dwell time for dust than the approximate age of the
bubble, although gas-grain friction is the primary mechanism of
dust-grain removal for grains of all sizes.

One possible explanation for the presence of dust in this environment
(that we will investigate in more detail in Section~\ref{proposedModel}) is
that dust grains could be the product of evaporating, colder, more
dense condensations, or `cloudlets', within the wind-blown bubble.
Perhaps dusty cloudlets are advected into the WBB, and are then
stripped away by the post-shock gas within the bubble.  This process
of cloudlet advection and destruction has, of course, been studied by
researchers before \citep[e.g.,][]{CowieMcKee1977, PittardEtAl2001a,
PittardEtAl2001b, Pittard2007} albeit not including the effects of
dusty cloudlets, to our knowledge.  Another possible source of dust
replenishment might be debris disks of low-mass stars formed along
with the O5V star in N49, which could be ejecting dust particles.

As dust emission seems to fit the observations, we continue to pursue
the role of dust in cooling the bubble in the following sub-sections.
We first show how the charge on dust varies within the bubble, and
then what effect the dust cooling has on the bubble; we will then
assemble these results into a model of N49.

\subsection{Charge on Dust in the Wind-Blown Bubble}

\begin{figure}[h!]
\begin{center}
\includegraphics[width=6cm,angle=90]{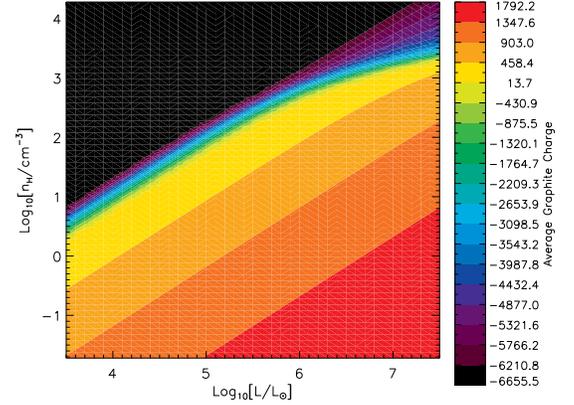}
\caption{The average graphite dust-grain charge in our Cloudy models
of the dusty wind-blown bubble.  The average here is taken over all
radii, and over all grain sizes, and is plotted vs. the luminosity of
the central star, $L$, and the density of the post-shocked gas,
$n_H$. (NB: this average is not weighted for the volume occupied by
grains at each radius, or for the relative populations of grains of
different sizes!)  As expected, at higher luminosities photo-emission
of electrons becomes an important process and the grains become
positively charged; on the ordinate, meanwhile, the grain charge
becomes more negative as $n_{\rm e}$
increases. \label{avgGraphiteCharge}}
\end{center}
\end{figure}

\begin{figure}[h!]
\begin{center}
\includegraphics[width=6cm,angle=90]{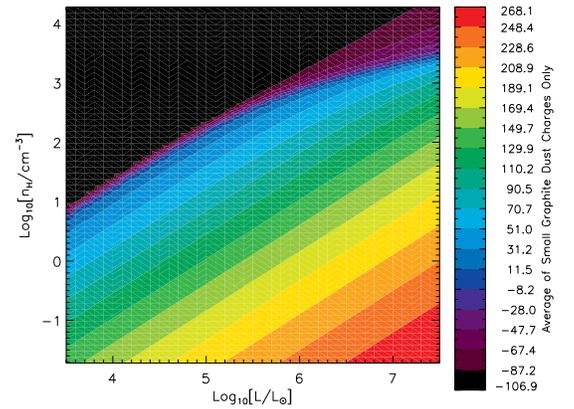}
\caption{As in Fig.~\ref{avgGraphiteCharge}, but with the average taken
only over the three smallest grain sizes ($a \la 10^{-2}\um$).  The
same trends can be seen as in Fig.~\ref{avgGraphiteCharge}, but for
the smaller grains, which dominate the grain distribution by number.
\label{avgSmallGraphiteCharge}}
\end{center}
\end{figure}

The impact of dust on the thermal balance of the dusty WBB is
determined largely by the charge on the grains.  In
Figures~\ref{avgGraphiteCharge} and \ref{avgSmallGraphiteCharge}, the
average graphite grain charges are presented: in
Figure~\ref{avgGraphiteCharge}, the average grain charge is computed
over all grain sizes and over all radii (with no weighting to take
into account the greater volume at larger radii, or the greater number
of smaller-diameter dust particles).
Figure~\ref{avgSmallGraphiteCharge} shows the average charge of grains
with radii $\la 0.01\um$.  We investigate these trends to better
understand the role of dust in WBBs.

There are two trends to note in Figures~\ref{avgGraphiteCharge} and
\ref{avgSmallGraphiteCharge}: first, as the luminosity increases, the
increased UV flux ejects more photo-electrons from the grains, pushing
the grains to more positive charges.  But, as the average ambient
density decreases, the ambient electron density decreases, leading to
fewer collisions with electrons, and resulting in more positively
charged grains.  Grain charge impacts both heating and cooling, which
are important in our simulations: collisional heating of grains
accounts for approximately 32\% of the grain heating, while radiative
heating of grains represents the remaining 68\% of grain heating.

In Figures~\ref{avgGraphiteCharge} and \ref{avgSmallGraphiteCharge},
we consider only graphite grains, but the average charge on silicate
grains is very similar to the graphite grains; the average
silicate-grain charge transitions to negative values at slightly
higher luminosity and lower density (by approximately 0.3 dex),
compared to the average graphite grain charge, but otherwise the
trends are very similar.

\subsection{What is the Effect of Dust on the Wind-Blown Bubble?}\label{modelResults}

\begin{figure}[ht!]
\begin{center}
\includegraphics[width=6cm,angle=-90]{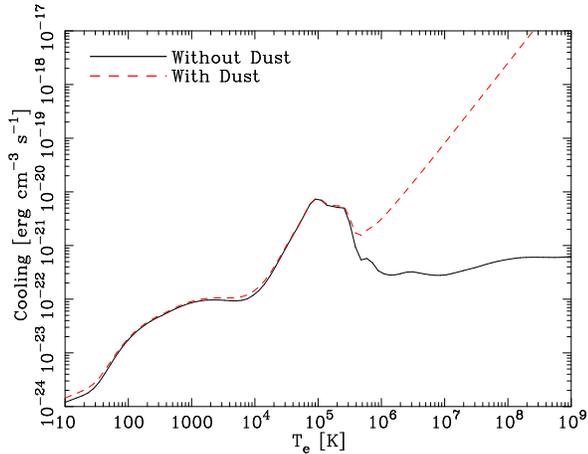}
\caption{Gas cooling curves within the WBB for the case where dust is
  not included (solid line) and when ISM dust is included (dashed line);
  the dust-to-gas mass fraction is set to $6.3\times10^{-3}$, as in
  all of our calculations.  As expected, where the atomic cooling
  starts to drop off near $4 \times 10^5$\,K, the dust cooling starts
  to dominate.  For the hot, post-shocked gas, if dust can survive in
  that region, dust will dominate the
  cooling.\label{coolingCurveComparison}}
\end{center}
\end{figure}

\begin{figure}[h!]
\begin{center}
\includegraphics[width=6cm,angle=90]{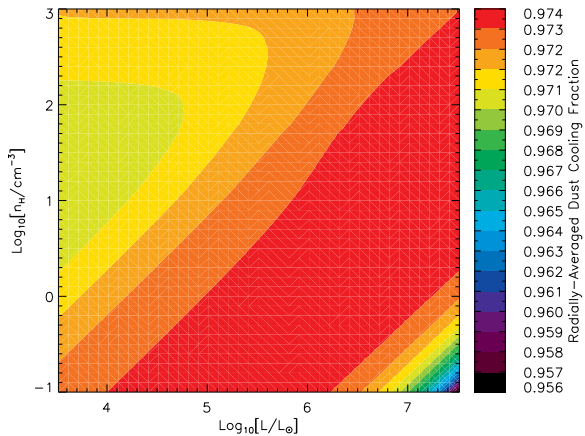}
\caption{The fraction of cooling from dust for a range of values of
  central stellar luminosity and hot-gas density (the parameters for
  the hot gas in N49, inferred from the mass outflow rate, are at the
  center of this plot).  The dominance of dust is evident across all
  of the parameter space surveyed; note the suppressed zero on the
  legend, at right. Only on the lower-right portion of the plot does
  Compton cooling start to become a factor. \label{dustCoolingFraction}}
\end{center}
\end{figure}

If dust can survive, dust cooling will be dominant in the hot,
post-shocked gas of a wind-blown bubble.  This is illustrated in
Figure~\ref{coolingCurveComparison}, where we compare cooling curves
from Cloudy where dust has not been included in the model (solid line)
and where dust has been included (dashed line).  We can see from
Figure~\ref{coolingCurveComparison} that in the hot, post-shocked gas
of the WBB \citep[see, e.g.,][]{FreyerEtAl2006}, dust cooling will be
dominant for $T \ga 5 \times 10^5$\,K.  We can see this for a wider
range of parameters in Figure~\ref{dustCoolingFraction}.

\begin{figure}[h!]
\begin{center}
\includegraphics[width=6cm,angle=-90]{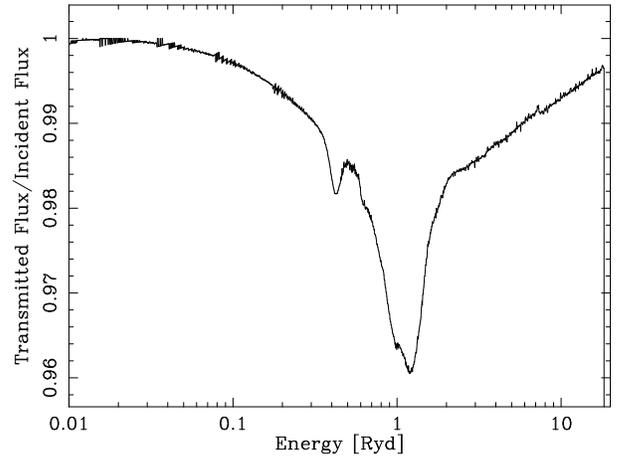}
\caption{The ratio of stellar flux transmitted through the wind-blown
  bubble to the incident stellar continuum.  The peaks in absorption
  correspond to increase in dust absorption properties at 2200\,\AA
  ($\sim 0.4$\,Ryd) and at $\sim 1$\,Ryd.  The cutoff in the fraction
  at approximately 20\,Ryd is due to the complete fall-off of stellar
  flux at that limit, so the ratio becomes ill-defined.
  \label{dustAbsorptionFraction}}
\end{center}
\end{figure}

But how does dust impact the WBB dynamics?  First, in considering the
size of the WBB, the dust column that we find is not large enough (in
the post-shocked region of the nebula) to absorb a significant
fraction of the ionizing radiation; overall, dust absorbs less than
3\% of the stellar flux in the energy range of 0.01\,Ryd to 20\,Ryd
(see Fig.~\ref{dustAbsorptionFraction}).
%This is not surprising: the
%column of dusty material in the bubble amounts to only $N_{\rm H} =
%2.4 \times 10^{19}$\,cm$^{-2}$; the small amount of absorption would
%be explained by an average dust absorption cross section in this
%regime of order $1 \times 10^{-21}$\,cm$^{-22}$/H~nucleon.  This is
%actually a bit high for the dust cross-section; I think the problem
%with this check is that there is also a significant amount of H
%absorption near 1 Ryd as well.
And yet, this small amount of dust can \textit{significantly} impact
the cooling and evolution of the WBB.

We present a simplified numerical model for N49 by integrating the
equations of motion for a \citet{WeaverEtAl1977} WBB and including
dust cooling.  For this, we use the standard equations of wind-blown
bubble expansion from \citet{WeaverEtAl1977} \citep[see, for a summary
of the equations,][]{HarperClarkMurray2009}, with the exception of a
slightly improved cooling term in the equation for the rate of
increase of mass inside the bubble:
\begin{equation}
\frac{dM_{\rm Bubble}}{dt} = C_1 T^{5/2}_{\rm Bubble} R^2_{\rm Bubble} (R_{\rm Bubble} -
R_{\rm Wind})^{-1} -
C_2 \frac{\mu L_{\rm Bubble}}{k T_{\rm Bubble}}
\end{equation}
where, instead of using $T_{\rm Bubble} = 2 \times 10^5$\,K as assumed
in \citet{WeaverEtAl1977}, we use the temperature in the wind-blown
bubble at each time-step, and calculate the luminosity, $L_{\rm b}$,
using the dust-dominated (at high $T$) cooling curve from Cloudy (note
that this equation assumes that the mass-loading from evaporation of
the surrounding dense shell is dominant; we will examine this
assumption at the end of Section~\ref{proposedModel}).  To start with
a simple model, and to isolate the effects of dust alone, we assume
that dust is added at the same rate as gas to the WBB, so that dust is
kept at a constant dust-mass-to-gas-mass ratio.

\begin{figure*}[h]
\begin{center}
\includegraphics[width=11cm,angle=-90]{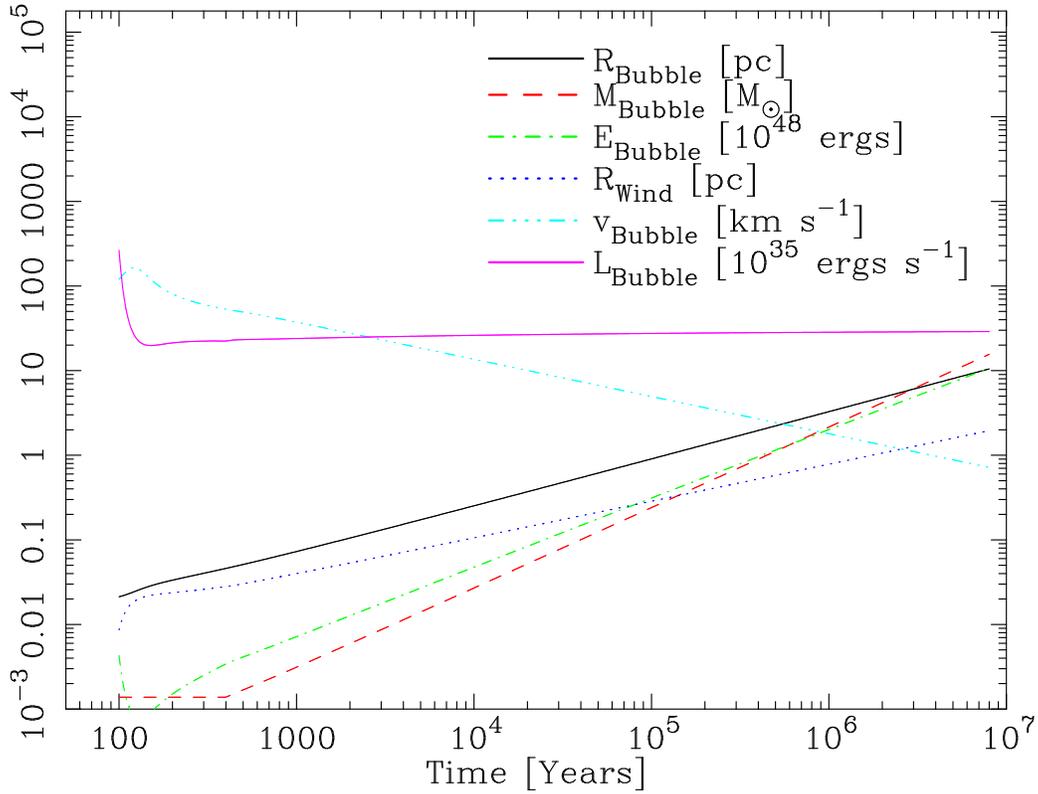}
\caption{Analytic calculation of the growth of a wind-blown bubble
  with dust cooling included. In the legend, $R_{\rm Bubble}$ refers
  to the outermost radius of the wind-blown bubble ($R_{\rm Bubble}$
  is labeled $R_2$ in Weaver et al.); in N49, this is identified with
  the innermost radius of the PAH emission.  $M_{\rm Bubble}$ shows
  the amount of mass that has been evaporated from the ISM shock and
  subsumed into the post-shocked gas ($M_{\rm Bubble}$ is $M_b$ in
  Weaver et al.).  $E_{\rm Bubble}$ is the total energy in hot gas
  within the post-shocked gas, $R_{\rm Wind}$ is the radius of the
  wind-dominated, dust-free region inside the post-shocked gas (these
  are $E_b$ and $R_1$ in Weaver et al., respectively).  $v_{\rm
  Bubble}$ is the velocity of the outermost shock of the bubble, and
  $L_{\rm Bubble}$ is the total luminosity (including dust emission
  and also including non-radiative cooling of electrons due to
  dust-grain charge) from the post-shocked gas (corresponding to $V_2$
  and $L_b$ in Weaver et al., respectively). The upturns in
  luminosity, energy density, and mass in the bubble at t=200 years
  are the transitions from \citet{WeaverEtAl1977} initial conditions
  to our dynamical model that includes dust cooling.  The
  near-constant luminosity is the result of the increasing size of the
  bubble compensating for both the slowly decreasing density in the
  bubble and the slow drop in cooling as the temperature decreases
  (this calculation assumes a constant temperature throughout the WBB
  at any given time, which slowly decreases over
  time). \label{simpleDynamicalModel}}
\end{center}
\end{figure*}

The result of this calculation is shown\footnote{We note that this
includes non-radiative cooling of electrons due to the electric
potential of the dust grains, and as such, we will overestimate the
observed luminosity of the bubble.}  in
Figure~\ref{simpleDynamicalModel}.  We have checked the calculation by
duplicating the results in \citet{WeaverEtAl1977} in the dust-free
case.  Note also the transition at $t \sim 10^2$\,years, in our
models, from dust-free initial conditions to the dusty WBB dynamics;
the initial radiation losses are high enough that $M_{\rm bubble}$
ceases to grow in mass, and for these times, we set $dM_{\rm
bubble}/dt = 0$.

\begin{figure*}[h]
\begin{center}
\includegraphics[width=11cm,angle=-90]{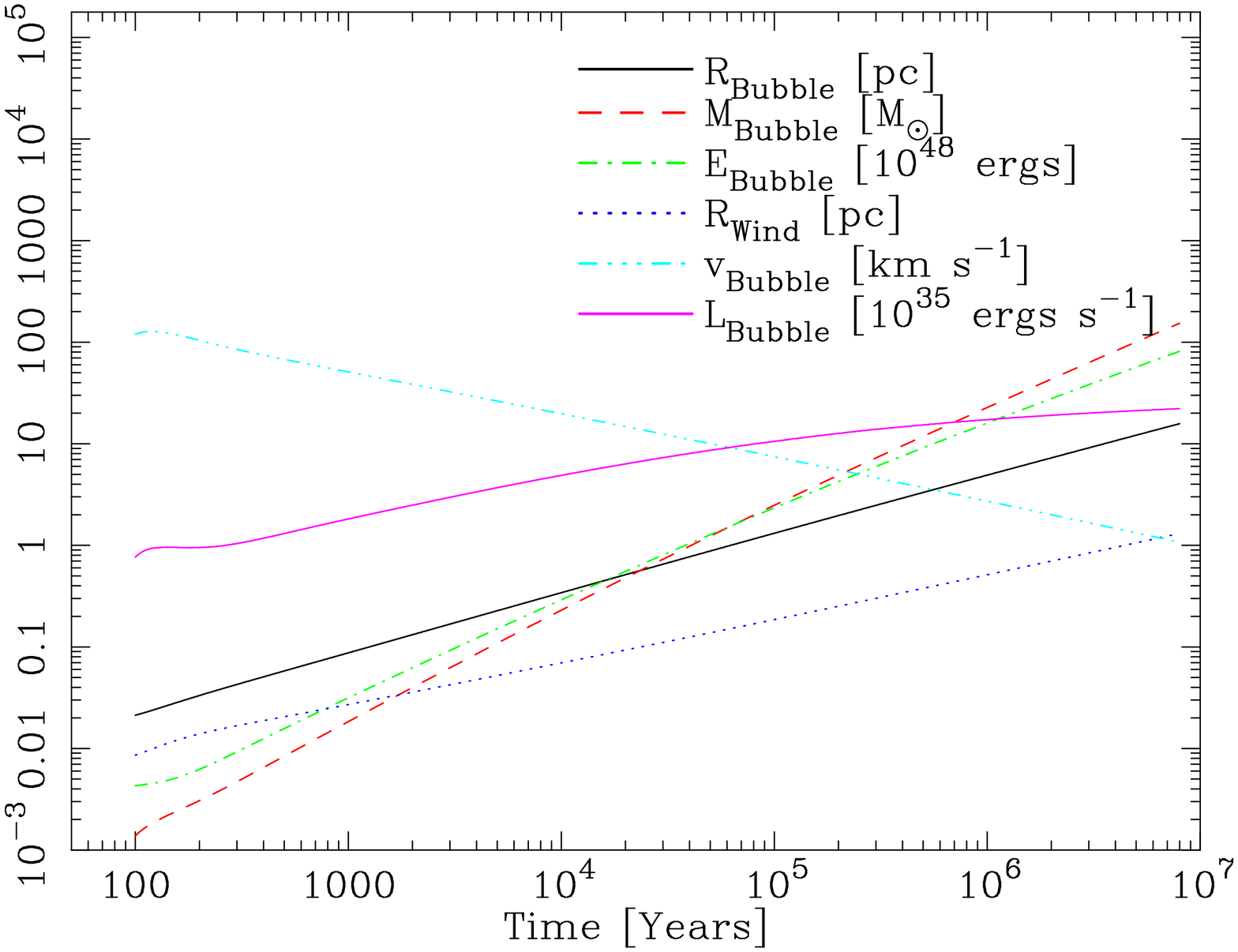}
\caption{As in Figure~\ref{simpleDynamicalModel} but \textit{without}
  dust cooling included. \label{simpleDynamicalModelNoDust}}
\end{center}
\end{figure*}

Figure~\ref{simpleDynamicalModelNoDust} shows the same model, with the
same initial parameters, except that dust is not included in this
model.  One can see that \citep[as in][]{WeaverEtAl1977}, the
luminosity increases with time, in contrast to the model with dust
cooling presented in Figure~\ref{simpleDynamicalModel}.  In fact, the
dusty WBB has a luminosity at $t = 10^4$~years that is approximately a
factor of six higher than the luminosity of the dust-free WBB.  As a
result, by an age of $10^7$~years, if the central star lives that
long, the energy contained within the dusty WBB is approximately a
factor of eight less than the energy in the dust-free bubble.

The temperature indicated by the dynamical model of this dusty
wind-blown bubble is approximately $3.5 \times 10^6$~K at an age of
$10^5$ years; this temperature only slightly decreases over time to
$\sim 3 \times 10^6$~K at $10^6$ years.  This model was therefore used
to set the fiducial temperature and density that we have used in
earlier sections of the paper, in order to be relatively
self-consistent with the parameters used.  (Earlier models of ours
assumed a temperature of $T \sim 10^7$\,K as in \citet{FreyerEtAl03};
as mentioned earlier, that different temperature does not affect our
results significantly.)

It is important to note that, in this model, the gas
temperature inside the dusty WBB is approximately the same as the
dust-free WBB for ages $\ga10^3$ years.  This is surprising, since
dust is such a strong coolant.  However, this coolant is most
important at early times, when the WBB temperature is $\sim 10^7$\,K;
at such high temperatures, the dust cooling rate is much higher than
the gas-derived cooling rate.  So, in the early evolution of the WBB,
when dust is introduced in the model, it is indeed a strong coolant:
the resultant WBB luminosity is very high, and the temperature and
energy in the bubble drop sharply.  In the Weaver et al. model, such a
high luminosity implies a lower energy input to the outer shell of the
WBB, and therefore a decrease in the evaporated mass-flux from that
shell into the bubble.  This decrease in mass flux and the energy lost
due to radiation in this $\sim10^7$\,K phase roughly track each other,
and so the overall temperature in the bubble does not change greatly.
Therefore, in this model, the dust impacts the growth of the bubble
(its size and mass), but does not greatly affect the temperature.

This WBB model also approximately reproduces other features of N49.
The inner radius of the bubble reaches 0.5~pc at an age of $3 \times
10^5$~years, and at the same time has an outer radius of approximately
1.8~pc, close to the 2.2~pc that we derive from observations.  The
expansion velocity of the bubble is approximately 3~km~s$^{-1}$ at
that same age.

But how can dust exist continuously in the post-shocked bubble?  We
consider that possibility in the next section.

\section{A Proposed Model for N49}\label{proposedModel}

\begin{figure*}[h]
\begin{center}
\includegraphics[width=12cm]{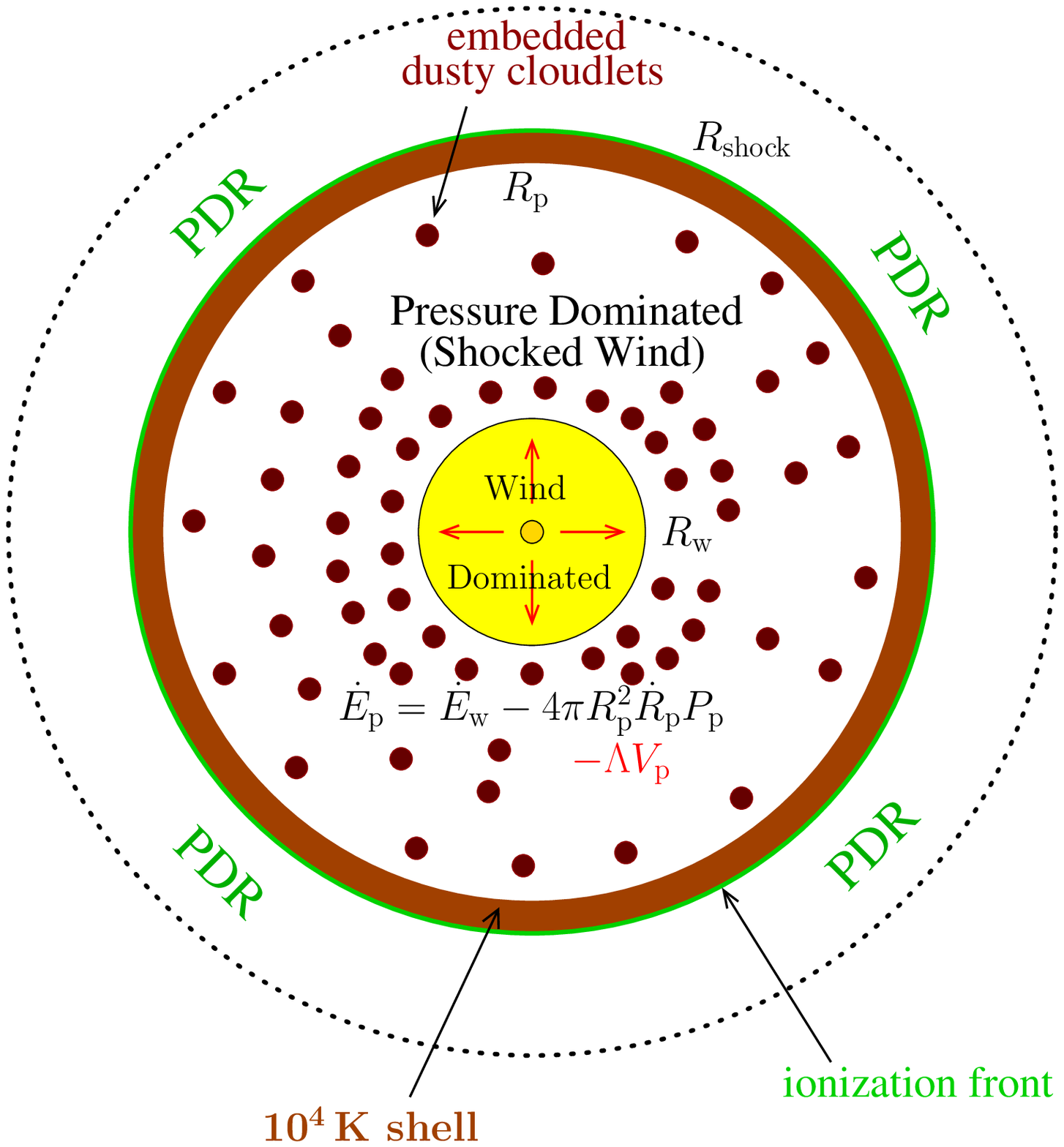}
\caption{Our schematic of the dusty, wind-blown bubble of N49.  The
  stellar wind has completely evacuated the central region of dust,
  leaving a dust-free, ``wind dominated'' zone of radius $R_{\rm w}$
  in the center.  We identify the observed central dip in
  azimuthally-averaged 24\,$\um$ emission in N49 with this
  wind-dominated region. Dust is driven from the post-shock gas by
  dust-gas friction and destroyed by sputtering on timescales about at
  least an order of magnitude less than the age of the wind-blown
  bubble; therefore, to resupply dust within the bubble, we
  hypothesize that small cloudlets are over-run by the expanding
  bubble, and slowly destroyed by heat conduction and
  ablation.\label{n49Schematic}}
\end{center}
\end{figure*}

Our photoionization and (as yet, quite simplified) dynamical modeling
of N49 shows the possible importance of dust on the bubble.  However,
as shown in earlier sections, due to the relatively small timescale of
dust advection in the post-shock bubble, that dust must be somehow
resupplied.  As mentioned in Section~\ref{dustSurvival}, we
hypothesize that small, dense ISM `cloudlets' have been over-run by
the ionization front of the bubble (see Figure~\ref{n49Schematic}).
When these cloudlets interact with the shocked wind, they are
gradually disrupted, releasing their dust and gas into the
post-shocked wind.  The dust grains from these cloudlets then yield
the 24$\um$ emission that is observed.  The dust adds to the radiative
cooling of the post-shock gas, and decreases the energy density in the
bubble over time ($\dot{E}_{\rm p}$ in Fig.~\ref{n49Schematic})
relative to dust-free wind-blown bubbles.

We note that, while the presence and importance of dust within
wind-blown nebula is a relatively new idea, the presence of
substructure, or `clumps', within nebulae has been inferred in a
variety of other observations, from early studies of line ratios in
the Orion Nebula \citep{OsterbrockFlather1959} to the impact of clumps
on observations of the radio-continuum spectral slope\footnote{The
radio-continuum spectral slope can also be modified by the power-law
drop-off of density with radius \citep{PanagiaFelli1975, Olnon1975}}
\citep{CassinelliHartmann1977, IgnaceChurchwell2004}.  Extensive
theoretical work has already been done on the inclusion of clumps and
the resulting mass-loading \citep[see][]{PittardEtAl2001a,
PittardEtAl2001b}; for a review, see \citet{Pittard2007} and
references therein.

To further check this picture, we estimate how quickly these cloudlets
are destroyed within the WBB, and check whether such resupply is
consistent with observations and the model.  To examine this, we
calculated the mass deposition from the cloudlets into the WBB, where
the cloudlets are stripped away by both ablation and evaporation due
to the fast-moving, hot post-shocked wind that surrounds them.  [The
ram pressure and thermal pressure of the post-shocked wind would,
together, suppress photoevaporation of the clouds; in addition, even
if photoevaporation were present and not suppressed, it would result
in mass-loss rates of order the ablation and conduction rates, so
would not significantly change our estimates, especially given other
uncertainties \citep{Pittard2007}.] To estimate the effects of these
processes, we used the approximate equations of \citet{Pittard2007}.
First, for ablation, where the external flow is subsonic, we
calculate:
\begin{equation}
\dot{M}_{\rm ablation} \approx \mathcal{M}^{4/3} (M_c c_c)^{2/3} (\rho
v)^{1/3} {\rm g~s}^{-1},\label{ablationOne}
\end{equation}
where $\mathcal{M}$ is the Mach number of the post-shocked wind flow,
$M_c$ is the mass of the cloudlet, $c_c$ is the speed of sound within
the cloudlet, $\rho$ is the mass density of the post-shocked wind, and
$v$ is the velocity of the post-shocked wind.  For a supersonic flow,
\begin{equation}
\dot{M}_{\rm ablation} \approx (M_c c_c)^{2/3} (\rho
v)^{1/3} {\rm g~s}^{-1}. \label{ablationTwo}
\end{equation}
Meanwhile, the mass loss to conduction is given by 
\begin{equation}
\dot{M}_{\rm conduction} \approx 2.75 \times 10^{19} \omega r_{\rm
c,pc} T^{5/2}_6 {\rm g~s}^{-1} \label{conductionOne}
\end{equation}
where $\omega=1$ gives the limit of classical evaporation (which we
adopt here), $T_6$ is the temperature in the wind-blown bubble in
units of $10^6$~K, and $r_{\rm c,pc}$ is the radius of the cloudlet in
parsecs \citep[again, see][]{Pittard2007}.

To use these equations, we take the density within the bubble from our
simple dynamical model (shown in Fig.~\ref{simpleDynamicalModel}) at
$t=5\times10^5$~years, and use the approximation for $v$ within the
wind-blown bubble from page 379 of \citet{WeaverEtAl1977}:
\begin{equation}
v \sim \frac{v_{\rm wind}}{4}\left( \frac{15}{16} \right)^{3/2}
\frac{R_{\rm wind}^2}{r^2},\label{weaverVel}
\end{equation}
again using the wind velocity, post-shocked wind temperature, and
bubble radius from our dynamical model.  We then check whether there
exist reasonable cloud parameters ($M_c$, $c_c$, and $r_{\rm c,pc}$)
that would yield the required mass-outflow rate given the above
ablation and conduction mass-loss estimates, and which would have
enough mass to survive for $5 \times 10^5$~years.

Specifically, we require that the cloudlets need to resupply the
observed dusty-gas mass ($\sim 2.5$\,M$_{\odot}$ of gas) every $10^4$
years (the shortest timescale for dust sublimation and evacuation from
the bubble).  If the wind-blown bubble is $\sim 5 \times 10^5$\,yrs
old, then 125\,M$_{\odot}$ of dusty gas is required to be resupplied
to the bubble interior in that time.  The average mass loss rate is
then, of course, $2.5 \times 10^{-4}$\,M$_{\odot}$\,yr$^{-1}$ from all
of the cloudlets embedded in the post-shock gas.  Can cloudlet
ablation or conduction (or both together?) in the post-shock region
match that rate, and with a reasonable number of clouds?

Looking at Equations \ref{ablationOne} through \ref{conductionOne},
the conductive and ablative mass-loss rates depend not only on the
number of cloudlets, but also the cloudlet density, radius, and
temperature.  With these unknown parameters, we cannot claim a unique
solution.  Of all of those parameters, we can only approximately set
the density within the cloudlets, choosing $n_{\rm cloudlet} =
10^5$\,cm$^{-3}$ as an approximate high-density component of the
external medium.  After a search of parameter space, we find that
requiring these clouds to lose mass at a rate of $\sim2.5 \times
10^{-4}$\,M$_{\odot}$\,yr$^{-1}$ can be satisfied if the clouds have a
radius of approximately 0.05 pc or $\sim 10^4$\,AU and a cloud
temperature of about $T \sim 100$\,K.  Interestingly, the size of
these clouds is in the range of clouds observed in
\citet{DeMarcoEtAl2006} and \citet{KoenigEtAl2008}.  The mass of each
individual cloud is approximately $1.3 M_{\odot}$; the mass of
dusty-gas within the bubble could be replenished by a few hundred of
these kinds of clouds, spread throughout the wind-blown bubble,
occupying only approximately 0.5\% of the volume of N49.

With these parameters, cloud ablation dominates for most of the
interior of the wind-blown bubble, whereas thermal evaporation due to
conduction is important only for $R \ga 1.5$~pc.  This transition
occurs because the velocity of the post-shocked gas decreases with
radius \citep[see Eq.~\ref{weaverVel}, from][]{WeaverEtAl1977} as
$r^{-2}$, so that ablation becomes less important with radius in the
wind-blown bubble.  If ablation were the only important cloud
destruction process, $\dot{M}_{\rm ablation} \propto \mathcal{M}^{4/3}
v^{1/3} = (v/c_{\rm s,ext})^{4/3} v^{1/3} \propto v^{5/3}$ (where
$c_{\rm s,ext}$ is the sound speed in the external medium), and since
$v \propto r^{-2}$, we would expect $\dot{M}_{\rm ablation}$ and the
resultant dust density to scale as $r^{-10/3}$ or approximately as
$r^{-3}$.  Conduction at larger radius adds another mass-loss term,
however, so we would expect the dust distribution to drop off somewhat
more slowly than $r^{-3}$.  Also, dust-gas friction will push dust to
larger radii within the bubble and would tend to flatten the radial
distribution of dust.  So, in the end, our model's assumed dust
distribution of $r^{-2}$ is not unreasonable.  Much more work is
needed to try to understand the details of this process however; we
have aimed merely to physically justify the assumptions of our model
and to try to build a reasonably self-consistent picture.  It must be
noted that cloud destruction in this environment is a complicated and
non-linear process, and so the above expressions are only rough
estimates of this process; the rate of cloud ablation would certainly
change over time as gas and dust are added to the post-shocked gas,
for instance.

One difficulty with the above simple explanation for dust-loading that
points to further work is the amount of dusty gas that is added to the
pressurized interior of the bubble.  We have employed the models of
\citet{WeaverEtAl1977}, but with 125\,$M_{\odot}$ of dense cloudlets
added to the bubble over $10^5$ years, the mass in the cloudlets will
overwhelm the addition of mass via thermal conduction from the outer
shell \citep[assumed to be the main component of mass addition
in][]{WeaverEtAl1977}, dominate mass addition to the interior of the
bubble, and cool the bubble interior and impact its dynamics
\citep[see,e.g.,][]{ArthurEtAl1993, PittardEtAl2001b, NazeEtAl2002}.
The necessity for so much added gas mass could point to improvements
that must be made to our initial assumptions and calculations: perhaps
such dense cloudlets have larger dust grains than we have assumed that
are not destroyed as quickly (but that yield smaller dust grains via
slow sputtering), or perhaps normal-sized ISM grains last longer than
the above estimates of dust destruction indicate.  The possibility of
longer-lived grains could actually be related to the added gas mass,
which could cool the bubble interior and impacts the gas dynamics
within the hot bubble; both of these processes may slow the
destruction of grains, although, possibly countering this, an
increased density within the bubble would (by itself) increase dust
sublimation rates proportionately.

Regardless, it is clear that the presence of dust and of cloudlets
point to the necessity of new and more detailed models of wind-blown
bubbles: when we run simple models with the above mass-injection rates
(from cloudlets), we find that the bubble quickly cools.  In fact,
even with mass-injection rates an order of magnitude lower than those
postulated above ($2.5\times10^{-5}$\,M$_{\odot}$\,yr$^{-1}$), the
hot-pressure bubble cools after only a few hundred years.  With mass
input rates of $2.5\times10^{-6}$\,M$_{\odot}$\,yr$^{-1}$, the bubble
expands for millions of years without cooling.  This could indicate a
critical upper limit for the addition of mass in cloudlets.  We stress
again, however, that the above estimates are very simple and point the
way to necessary improvements of the basic models of wind-blown bubble
expansion, possibly along the lines of models such as those of
\citet{HanamiSakashita1987} and \citet{Pittard2007}, or the models of
cloudlets within supernova remnants, as in \citet{PittardEtAl2003}.

\section{Conclusions}\label{conclusions}

We were intrigued by the possibility of dust within the wind-blown
bubble 'N49' \citep{WatsonEtAl2008}.  Our goal was to explain the
present observations with dust, and to try to understand how dust may
survive within that environment.  In this paper, we found that both
the processes of dust-gas friction and sputtering of small dust grains
can lead to the evacuation of dust grains from WBBs on relatively
short timescales of approximately $10^4$~years.  We also found,
though, that if dust is present, it has a very significant effect on
the structure of the wind-blown bubble: decreasing the energy within
the bubble due to dust cooling, and therefore decreasing the size of
the bubble relative to dust-free bubbles.  However, in order for dust
to be present in N49 today, that dust must be replenished; we
hypothesize that high-density cloudlets ($n_{\rm c} \sim
10^5$\,cm$^{-3}$) can be overrun and subsumed by the wind-blown
bubble.  Those cloudlets are then gradually destroyed by ablation and
evaporation in the post-shock wind in the bubble, but on timescales
long enough that the cloudlets would continue to supply the required
density of dust to N49 over at least $5 \times 10^5$~years.  Such
substructure has already been suggested by other researchers
\citep[e.g.,][]{OsterbrockFlather1959, Pittard2007}; what is new here
is the consideration of dust.  We caution that the dust may be
supplied by other processes, too; for instance, the ejection of
dust-grains from the debris disks of nearby low-mass Young Stellar
Objects might also be important.

This model has several implications.  First, the external density
around N49 must be fairly high to support high-density clouds ($n_{\rm
c} \sim 10^5$~cm$^{-3}$); our dynamical model also assumes $n_{\rm
external} \sim 10^4$\,cm$^{-3}$ in the external medium.  This should
be tested by constraining the density through ammonia observations
(Cyganowski et al., in preparation).  Also, as dusty bubbles age, if
the cloudlet density is constant with radius within the bubble, dust
should gradually be evaporated from the inside out, such that there
should be wind-blown bubbles with only an outer-rim of 24$\um$
emission; this could be tested by comparing free-free and 24$\um$
observations of WBBs.  Finally, a similar structure of 24$\um$
emission should be visible in other wind-blown bubbles; as long the
bubble is young enough (ages of less than $10^6$~years for the cloud
parameters outlined herein) that it has not destroyed all of the dusty
cloudlets, similar 24$\um$ emission should be seen; there are hints of
this already (C. Watson, personal communication).  Finally, the large
amount of mass-loading that appears to be necessary to maintain dust
in the WBB may lead to large amounts of cooling and a short lifetime
for the \citet{WeaverEtAl1977} pressure-driven phase.

\acknowledgments

This work was supported by NSF AST-0808119, NASA/\textit{Spitzer}
Grant 1317498, NSF AST-0507367, NSF AST-0907837 and NSF PHY-0215581 \&
NSF PHY-0821899 (to the Center for Magnetic Self-Organization in
Laboratory and Astrophysical Plasmas).  This research has made use of
NASA's Astrophysics Data System.  The authors also wish to thank Gary
Ferland, Peter van Hoof, and all of their collaborators for their work
on the photoionization code Cloudy.  The authors thank Barb Whitney,
Kenny Wood, and Ellen Zweibel for helpful conversations; many thanks
also to Marilyn Meade for supplying MIPS images ``to order.''
Finally, we thank the referee for their time and effort in reviewing
the paper and asking detailed, insightful questions.

\bibliography{ms}

\end{document}